\documentclass[conference]{IEEEtran}
\IEEEoverridecommandlockouts
\usepackage{cite}
\usepackage{amsmath,amssymb,amsfonts}
\usepackage{graphicx}
\usepackage{textcomp}
\usepackage{xcolor}
\usepackage{graphicx}
\usepackage{subfigure}
\usepackage{amssymb,amsmath}
\usepackage{textcomp}
\usepackage{cite}
\usepackage{float}
\usepackage{bm}
\usepackage{here}
\usepackage{soul, color}
\usepackage{multirow}
\usepackage{booktabs}
\usepackage{multicol}
\usepackage{verbatim}
\usepackage{svg} 
\usepackage{graphicx}
\usepackage{CJKutf8}
\usepackage{amsmath}
\usepackage{algorithm}
\usepackage{algpseudocode}

\usepackage{amsmath}
\usepackage{amssymb,amsmath}
\usepackage{textcomp}
\usepackage{cite}
\usepackage{float}
\usepackage{bm}
\usepackage{here}
\usepackage{soul, color}
\usepackage{multirow}
\usepackage{booktabs}
\usepackage{multicol} 
\usepackage{array}

\usepackage{graphicx}
\usepackage{caption}

\def\BibTeX{{\rm B\kern-.05em{\sc i\kern-.025em b}\kern-.08em
    T\kern-.1667em\lower.7ex\hbox{E}\kern-.125emX}}
\title{Class E/EF Inductive Power Transfer to Achieve Stable Output under Variable Low Coupling}

\begin{document}

\author{
    \IEEEauthorblockN{Yifan Zhao$^{1}$, Mowei Lu$^{1}$, Heyuan Li$^{2}$, Zhenbin Zhang$^3$, Minfan Fu$^2$ and Stefan M. Goetz$^1$*}
    \IEEEauthorblockA{\textit{$^1$ Department of Engineering, University of Cambridge, Cambridge, United Kingdom.}}
    \IEEEauthorblockA{\textit{$^2$ School of Information Science and Technology, ShanghaiTech University, Shanghai, China.}}
    \IEEEauthorblockA{\textit{$^3$ School of Electrical Engineering, Shandong University, Jinan, China.}}
    \thanks{This work was supported by National Natural Science Foundation of China under Grant 52477013 and Lingang Laboratory under Grant NO. LG-GG-202402-06-10.}
    \IEEEauthorblockA{\textit{Email: smg84@cam.ac.uk}}
}
\maketitle

\begin{abstract}
This paper develops an  inductive power transfer (IPT) system with stable output power based on a Class E/EF inverter. Load-independent design of Class E/EF inverter has recently attracted widespread interest. However, applying this design to IPT systems has proven challenging when the coupling coefficient is weak. To solve this issue, this paper uses an expanded impedance model and substitutes the secondary side's perfect resonance with a detuned design. Therefore, the system can maintain stable output even under a low coupling coefficient. A 400 kHz experimental prototype validates these findings. The experimental results indicate that the output power fluctuation remains within 15\% as the coupling coefficient varies from 0.04 to 0.07. The peak power efficiency achieving 91\%.
\end{abstract}

\section{Introduction}

In recent years, inductive wireless power transfer technology has conquered various fields and demonstrated its versatility and applicability across diverse sectors. Unlike conventional charging methods that offer robust wired connections and generally reliable power transfer~\cite{ipt,ipt3,huiwang3,xly2,xx,nss,wzq,mowei2,mowei3,2,3,4,5,6,9,10,11,7,8}, inductive power transfer (IPT) systems are inherently sensitive to coupling, which means that any variations in the coupling coefficient between the transmitter and the receiver can cause significant fluctuation in the output power. The power fluctuation can impact the efficiency and stability of the power transfer process. Consequently, the establishment of an IPT system that is robust against coupling variations, capable of maintaining consistent and reliable power transfer despite changes in coupling conditions, is crucial for advancing inductive power transfer technology and ensuring its widespread adoption and success in various applications~\cite{leigu1,leigu2,tianunderwater,minjie1,tian2024design,zhao2023detuned,ss,xtype+s}.

The single switch Class E / EF inverter has attracted widespread application across various industries and applications due to its exceptional cost-effectiveness and high operational efficiency~\cite{rivas1,rivas2,rivas3,mingliu1}. However, the intrinsic load sensitivity of resonant converters poses significant challenges when feeding wireless power transfer systems. Consequently, a load-independent design of the class E / EF inverter seems to constitute a promising solution to overcome the challenge~\cite{pauld}. However, the implementation of a load-independent Class E / EF inverter is restricted by a minimum resistance requirement~\cite{pauld}. This means that the inverter must be connected to a load with a resistance that is above a certain threshold to ensure stable and efficient operation. When deployed within an IPT system, based on the reflected impedance model, there is an inherent lower bound on the coupling coefficient that must be met to maintain stable power. Given that the majority of IPT systems operate under weak coupling, this seemingly optimal scheme harbors substantial limitations.

This study comprehensively analyses and solves the critical failure of load-independent design in Class E/EF driven IPT systems when they are operating under weak coupling conditions. In response, we propose an innovative solution by substituting the ideal resonance on the secondary side with a detuned design. This detuned design incorporates an expanded impedance model~\cite{budengshi,zukangliang} to further explore the potential. This paper found that when the reflection impedance shifts from resistive to capacitive, it can effectively counteract the destabilizing effects of weak coupling. Moreover, this paper further elucidate theoretically why an inductive nature is more favorable for stable output in IPT systems.

\section{Topology of Class-E/EF Inverter Based IPT System}

\begin{figure}[htbp]
		\centering
		\subfigure[]{\includegraphics[width=0.9\linewidth]{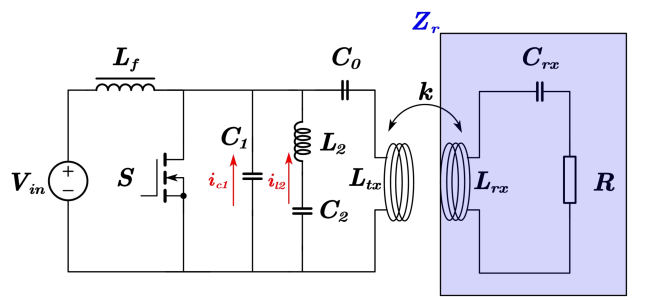}}
		\subfigure[]{\includegraphics[width=0.9\linewidth]{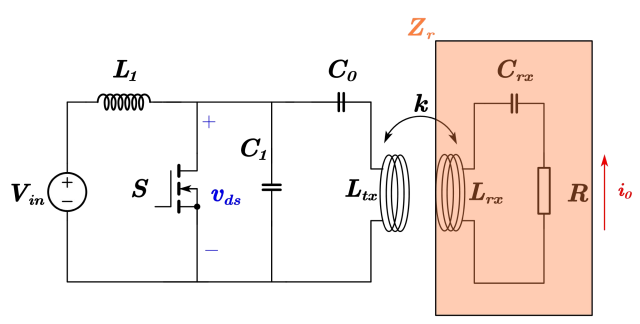}}
		\caption{Circuit model. (a) Basic IPT system driven by class EF inverter. (b)  Basic IPT system driven by class E inverter.}
		\label{Fig_Circuit model}
	\end{figure}

Fig~\ref{Fig_Circuit model}(a) depicts the topology of a basic IPT system driven by a Class E inverter. \( V_{dc} \) is the input voltage. \( S \) is the switch, whose duty cycle is \( D \) and frequency is \( f_{s} \). The TX coil's self inductance \( L_{tx} \) would resonate with \( C_{0} \) at frequency \( f_{s} \) with additional reactance \( X \). Please note that \( L_{tx} \) will resonate with \( C_{0} \) directly, which is different from a traditional resonate tank. In contrast to the traditional Class E inverter, \( L_{1} \) serves as a resonant inductor in place of a choke. The resonance frequency between \( L_{1} \) and \( C_{1} \) could be normalized with respect to \( w_{s} \); a frequency factor \( q \) is then defined as
\begin{equation}
X=\omega_{s}L_{tx}-1/(\omega_{s}C_{0}),
\end{equation}
\begin{equation}
    q=1/\left(\omega_{s}\sqrt{L_{1}C_{1}}\right).
\end{equation}

Meanwhile, Fig.~\ref{Fig_Circuit model}(b) illustrates the topology of the IPT system driven by the Class EF inverter, where the input inductor \( L_{f} \) serves as an RF choke. \( C_{1} \) is a shunt capacitor that absorbs the switch junction capacitance. At \( f_{s} \), \( L_{0} \) and \( C_{0} \) resonate with the additional reactance \( X_{0} \). The shunt tank \( L_{2} \) and \( C_{2} \) would resonate at \( f_{2} \), which can be described as
\begin{equation}
X_{0}=\omega_{s}L_{tx}-1/(\omega_{s},C_{0})
    \label{eq2}
\end{equation}
\begin{equation}
f_2=1/\left(2\pi\sqrt{L_2C_2}\right).
\end{equation}

\section{Design Methodology}

\subsection{Load-independent Class-E inverter}

When the quality factor \( Q \) is high, the output current \( i_{0} \) and voltage across the switch \( v_{ds} \) can obtained as follows with $\beta$ and \( p \) as given:
\begin{equation}
 i_{o}(\omega t)=I_{m}\sin(\omega t+\phi)
 \label{i0}
\end{equation}
\begin{equation}
\begin{aligned}
\frac{\nu_{\mathrm{DS}}(\omega t)}{V_{\mathrm{in}}}=\frac{I_{m}}{\omega C_{1}V_{\mathrm{in}}}\int_{2\pi D}^{\omega t}\frac{i_{C_{1}}(\tau)}{I_{m}}\, d\tau=q^{2} p \beta(\omega t)
\end{aligned}
\label{vds}
\end{equation}
\begin{equation}
    \beta(\omega t)=\int_{2\pi D}^{\omega t}\frac{i_{C_{1}}(\omega t)}{I_{m}}\, d\omega t,p=\frac{\omega L_{1}I_{m}}{V_{\mathrm{in}}}
\end{equation}

The switch voltage is filtered by the total impedance of \( j_{X} \) and \( Z_{r} \). Consequently, upon applying the Fourier transform to \( v_{ds} \), its sine and cosine components are respectively applied to the resistive and reactive parts of the entire branch, i.e.,
\begin{equation}
    \begin{aligned}
&\frac{\nu_{Z_{r}}}{V_{\mathrm{in}}}=\frac{1}{\pi}\int_{2\pi D}^{2\pi}\frac{\nu_{\mathrm{ds}}(\omega t)}{V_{\mathrm{in}}}\sin(\omega t+\phi)d\omega t \\
&=\frac{q^{2}p}{\pi}\int_{2\pi D}^{2\pi}\beta(\omega t)\sin(\omega t+\phi)\, d\omega t=\frac{q^{2}p}{\pi}\psi_{1},
\end{aligned}
\end{equation}
\begin{equation}
 \begin{aligned}
    &\frac{\nu_{jX}}{V_{\mathrm{in}}}=\frac{1}{\pi}\int_{2\pi D}^{2\pi}\frac{\nu_{\mathrm{ds}}(\omega t)}{V_{\mathrm{in}}}\cos(\omega t+\phi)d\omega t\\&=\frac{q^{2}p}{\pi}\int_{2\pi D}^{2\pi}\beta(\omega t)\cos(\omega t+\phi)d\omega t=\frac{q^{2}p}{\pi}\psi_{2}.
\end{aligned}
\end{equation}

There are two design objectives: Constant ZVS and constant output, which can be derived as
\begin{equation}
\begin{aligned}
    &\frac{\partial}{\partial p}\Bigg(\frac{q^{2}p}{\pi}\psi_{1}\Bigg)=\frac{q^{2}}{\pi}\frac{\partial p\psi_{1}}{\partial p}=0 \mathrm{~over~range~of~} p,\\
\end{aligned}
\end{equation}
\begin{equation}
\beta(2\pi)=0\mathrm{~over~} \mathbb{D}_{_p}.
 \end{equation}

Therefore, the design variable can be obtained as  
\begin{equation}
\begin{aligned}
    &\frac{\nu_{jX}}{V_{\mathrm{in}}}=\frac{q^{2}p}{\pi}\psi_{2}(q,p,\phi,D)=\frac{i_{m}X}{V_{\mathrm{in}}}=p \frac{X}{\omega L_{1}} ,\\
    \end{aligned}
\end{equation}

\begin{equation}
\frac{X(q,\phi,D)}{\omega L_1}=\frac{q^2}{\pi}\psi_2(q,\phi,D)
\end{equation}

The solution of above equation is \textit{q}=1.2915, $\frac{X(q,\phi,D)}{\omega L_{1}}=0.2663$.

\subsection{Load-independent Class-EF inverter}

Beginning with the series  \( L_{2} \)\,{}\( C_{2} \) network, its current can obtained as 
\begin{equation}
\begin{aligned}
    &\frac{i_{L_{2}}}{I_{\mathrm{in}}}(\omega t)=A_{2}\cos\left(q_{2}\omega t\right)+B_{2},
    \end{aligned}
\end{equation}
\begin{equation}
\sin\left(q_{2}\omega t\right)-\frac{q_{2}^{2}p}{q_{2}^{2}-1}\sin(\omega t+\phi)+\frac{1}{k+1},
\end{equation}
\begin{equation}
\begin{aligned}
    &k=\frac{C_{1}}{C_{2}},q_{2}=\frac{1}{\omega}\sqrt{\frac{C_{1}+C_{2}}{L_{2}C_{1}C_{2}}}=q_{1}\sqrt{\frac{k+1}{k}},\\
    \end{aligned}
    \label{k}
\end{equation}
\begin{equation}
p =\frac{C_{2}}{C_{1}+C_{2}}\frac{I_{m}}{I_{\mathrm{in}}}=\frac{1}{k+1}\frac{I_{m}}{I_{\mathrm{in}}},
\label{p}
\end{equation}
where  \( k \),  \( q_{2} \),  \( p \) can be obtained from~\eqref{k} and \eqref{p}.

Meanwhile, the current flow into \( C_{1} \) and the drain voltage follow
\begin{equation}
    \frac{i_{C_{1}}}{I_{\mathrm{in}}}(\omega t)=1-p(k+1)\sin(\omega t+\phi)-\frac{i_{L_{2}}}{I_{\mathrm{in}}}(\omega t),
\end{equation}
\begin{equation}
 \frac{\nu_{\mathrm{ds}}(\omega t)}{V_{\mathrm{in}}}=2\pi\frac{\beta(\omega t)}{\alpha},
\end{equation}
\begin{equation}
\begin{aligned}
    &\beta(\omega t)=\int_{2\pi D}^{\omega t}\frac{i_{C_{1}}}{I_{\mathrm{in}}}(\tau)d\tau,\\
    \end{aligned}
\end{equation}
\begin{equation}
    \alpha=\int_{2 \pi D}^{2\pi}\beta(\omega t)d\omega t.
\end{equation}
Similar to the Class-E design, the Fourier transform yields
\begin{equation}
\begin{aligned}
    &\frac{\nu_{Z_{r}}}{V_{\mathrm{in}}}=\frac{2}{\alpha}\int_{2\pi D}^{2\pi}\beta(\omega t)\sin(\omega t+\phi)d\omega t=\frac{2}{\alpha}\psi_{1},\\
    &\frac{\nu_{jX}}{V_{\mathrm{in}}}=\frac{2}{\alpha}\int_{2\pi D}^{2\pi}\beta(\omega t)\cos(\omega t+\phi)d\omega t=\frac{2}{\alpha}\psi_{2}.
    \end{aligned}
\end{equation}

There are also two design objectives: Constant ZVS and constant output power per
\begin{equation}
    \frac{\partial}{\partial p}\Bigg(p\frac{\psi_{1}(p)}{\alpha(p)}\Bigg)=0 \quad
    \mathrm{over}~ \mathbb{D}_{p},
\end{equation}
\begin{equation}
\beta(2\pi)=0 \quad \mathrm{over}~ \mathbb{D}_{_p}.
\end{equation}

Therefore, the design variable can be obtained as
\begin{equation}
 \frac{1}{\omega R_{L}C_{1}}=\frac{\pi p^{2} (k+1)^{2}}{\alpha(p)},
 \label{1}
\end{equation}
\begin{equation}
\omega XC_{1}=\frac{1}{\pi p(k+1)}\psi_{2}\big(q_{1}, k,\phi,D\big),
\label{2}
\end{equation}
\begin{equation}
\I_{m}=2\frac{\psi_{1}(p)}{\alpha(p)}\frac{V_{\mathrm{in}}}{R}.
\end{equation}

The solution of above equation is \( q \)=1.3, $\frac{X(q,\phi,D)}{\omega L_{1}}=0.3533$.

\subsection{Detuned Compensation Design}

As mentioned before, this design has the minimum coupling-variation requirement. In scenarios with low coupling coefficients, the detuned secondary design allows for the full release of design variables. 
\subsubsection{Expanded impedance model}

To employ the expanded impedance model, the initial step involves representing circuit elements through their respective impedance matrices. Define the critical harmonic order range as $[-N, N]$, where harmonics outside this range are disregarded. It is crucial to set this range to optimize computational efficiency without compromising design accuracy. Consequently, all impedance expressions form square matrices of dimension $(2 N+1)$. Notably, resistors, inductors, and capacitors are represented by diagonal matrices.

The matrix form of a resistor follows
\begin{equation}
\mathbf{Z}_{\mathbf{R}}=\left[\begin{array}{ccccc}
R & \cdots & 0 & \cdots & 0 \\
\vdots & \ddots & \ddots & \ddots & \vdots \\
0 & \ddots & R & \ddots & 0 \\
\vdots \\
\vdots & \ddots & \ddots & \ddots & \vdots \\
0 & \cdots & 0 & \cdots & R
\end{array}\right],
\label{zr}
\end{equation}
the matrix form of inductors equivalently
\begin{equation}
\mathbf{Z}_{\mathbf{L}}=\left[\begin{array}{ccccc}
-j N \omega L & \cdots & 0 & \cdots & 0 \\
\vdots & \ddots & \ddots & \ddots & \vdots \\
0 & \ddots & 0 & \ddots & 0 \\
\vdots & \ddots & \ddots & \ddots & \vdots \\
0 & \cdots & 0 & \cdots & j N \omega L
\end{array}\right].
\label{zl}
\end{equation}

The matrix form of capacitors follows
\begin{equation}
\mathbf{Z}_{\mathbf{C}}=\left[\begin{array}{ccccc}
-j N \omega C & \cdots & 0 & \cdots & 0 \\
\vdots & \ddots & \ddots & \ddots & \vdots \\
0 & \ddots & 0 & \ddots & 0 \\
\vdots & \ddots & \ddots & \ddots & \vdots \\
0 & \cdots & 0 & \cdots & j N \omega C
\end{array}\right]^{-1}.
\label{zc}
\end{equation}

The transistor is modeled as a resistance that varies with time, featuring a significant OFF-state resistance \( R_{\text{off}} \) and a minor ON-state resistance \( R_{\text{on}} \). This applies to all harmonic orders \( p \) within the range \([-N, N]\).

\begin{equation}
R_{S, p}= \begin{cases}R_{\mathrm{ON}} D+R_{\mathrm{OFF}}(1-D), & p=0 \\ R_{\mathrm{ON}}-R_{\mathrm{OFF}} \frac{\sin (p \pi D)}{p \pi} e^{-j p \pi D}, & p \neq 0\end{cases}.
\end{equation}

Note that \( D \) is the duty cycle of the driving signal. The impedance expression of the active switch is
\begin{equation}
\mathbf{Z}_{\mathrm{S}}=\left[\begin{array}{ccccc}
R_{S, 0} & \cdots & R_{S,-N} & \cdots & R_{S,-2 N} \\
\vdots & \ddots & \ddots & \ddots & \vdots \\
R_{S, N} & \ddots & R_{S, 0} & \ddots & R_{S,-N} \\
\vdots & \ddots & \ddots & \ddots & \vdots \\
R_{S, 2 N} & \cdots & R_{S, N} & \cdots & R_{S, 0}
\end{array}\right].
\end{equation}

The dc input voltage can be expressed as
\begin{equation}
\mathbf{V}_{in}=\left[\begin{array}{lllllll}
0 & \cdots & 0 & V_{\text {in }} & 0 & \cdots & 0
\end{array}\right]^{\mathrm{T}} .
\end{equation}

Besides simulating the steady-state waveforms, the supplied power and load power at steady state can also be derived as
\begin{equation}
\begin{aligned}
\bar{P}_{\text {in }} & =\frac{1}{T} \int_T\left|V_{\mathrm{DC}}(t) i_0(t)\right| \mathrm{d} t \\
& =\sum_{k=-\infty}^{+\infty} V_{\mathrm{DC}, k} I_{0, k} \\
& =\mathbf{V}_{\mathbf{D C}}^{\mathrm{T}} \mathbf{Y}_{\mathbf{C l a s s}-\mathbf{E}} \mathbf{V}_{\mathbf{D C}}, \\
\bar{P}_{\text {out }} & =\mathbf{V}_{\mathbf{C}}^{\mathrm{T}} \mathbf{Y}_{\text {load }} \mathbf{V}_{\mathbf{C}}.
\end{aligned}
\end{equation}
\subsubsection{Analysis Results}

\begin{figure}[htbp]
    \centering
    \includegraphics[width=\linewidth]{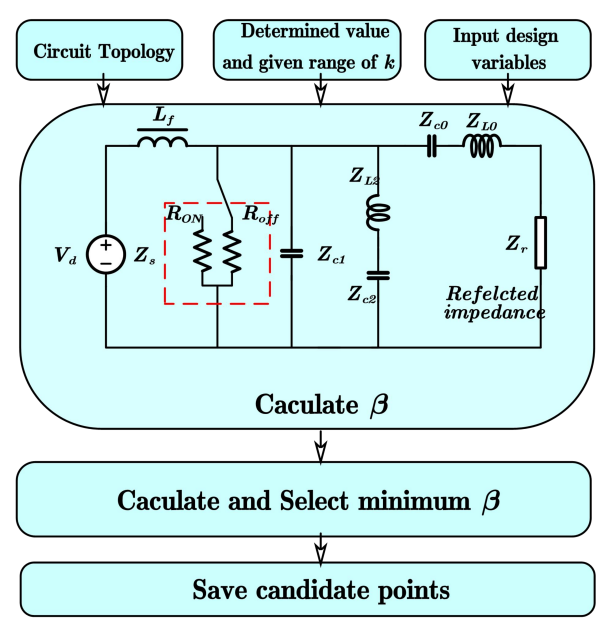}
    \caption{Design flow chat}
    \label{fig2}
\end{figure}

As depicted in Fig.~\ref{fig2}, this work uses an expanded impedance model. For the specific topology, determined values, and given range of \( k \), we can calculate the power drop ratio $\beta$, and by select minimum $\beta$, can get the candidate points, which are listed in TABLE I.

\begin{table}[htbp]
		\centering
		\caption{{Parameters and constraint condtions of design.}}
		\centering \begin{tabular}{c|c|c|c}
			\toprule
			Parameters & Value & Parameters & Value  \\
			\midrule
			$L_1$ & 10 $\mu \mathrm{H}$ & $V_{in}$ & 30 V 	\\
			$L_{tx}$ & 140 $\mu \mathrm{H}$ & $L_{rx}$ & 50 $\mu \mathrm{H}  $	\\
			$R_L$	& 12.5 $\Omega$ & $C_0$ & 1.15 nF     \\
			$C_1$ & 9.49 nF & $C_{rx}$ & 3.3 nF\\
			$Q_{tx}$ & 350 & $Q_{rx}$ & 251\\
			\bottomrule
		\end{tabular}%
		\label{tab:Constraint}%
	\end{table}%

The selected design point reveals an inductive characteristic on the secondary side. This occurs because an inductive secondary side results in a capacitive impedance when reflected to the primary side. Within a load-independent design framework, even in scenarios where low coupling leads to performance degradation, the voltage at the switching node remains relatively constant. Consequently, if the reflected impedance to the primary side is capacitive, it can compensate for a portion of the transmitter coil's impedance and secure a larger share of the voltage, thereby facilitating the maintenance of a stable output.

\section{Simulation Results}
The point selection model does not account for the effects of parasitic resistances and the nonlinear junction capacitance. Therefore, additional validation through simulation is necessary to confirm the model’s effectiveness. Using the parameters of Table I, the quality factors of the inductive components are taken into consideration, and the switch is modeled with a Spice model with the GaN transistor GS66508B.

\begin{figure}[htbp]
    \centering
    \includegraphics[width=1.02\linewidth]{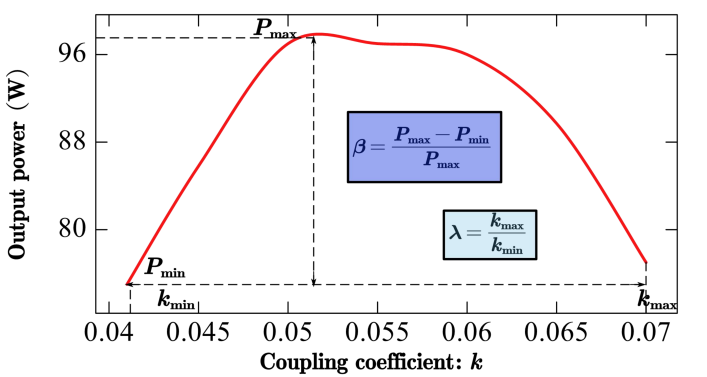}
    \caption{\( k\)-dependent power under  proposed design.}
    \label{fig3}
\end{figure}

Fig.~\ref{fig3} illustrates the relationship between output power and varying \( k\). It demonstrates that for \( k<0.05 \), \( P_{0} \) increases with \( k\), whereas for \( k>0.05 \), \( P_{0} \) decreases with \( k \). In this context, \( \beta \) is 20\%. 

\begin{figure}[htbp]
    \centering
    \includegraphics[width=\linewidth]{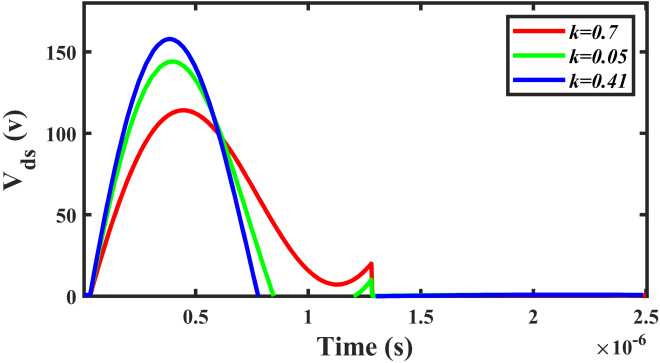}
    \caption{\( v_{ds} \) under different \( k\).}
    \label{fig4}
\end{figure}

Fig.~\ref{fig4} presents the waveform of the voltage across the switch \( v_{ds} \). It can be observed that under this design, ZVS only holds true for certain coupling coefficients. However, this does not necessarily imply an ineffective design, as the core objective of this paper is to achieve stable output, and constant ZVS is not the primary design goal. On the other hand, the implementation of ZVS aims to achieve high efficiency. Section V describes the experimental verification to demonstrate that this design, even with partial loss of ZVS, can still maintain high efficiency.

\section{Experimental Results}
An experimental IPT system, designed as previously described, is depicted in Fig.~\ref{fig5}. Both the TX and RX coils are crafted from Litz wire, with specifications of 0.1 mm by 150 strands and a diameter of 1.7 mm. All system parameters are detailed in Table I. The circuit uses high-precision, high-quality-factor NP0/C0G ceramic capacitors. 
\begin{figure*}[htbp]
    \centering
    \includegraphics[width=0.9\linewidth]{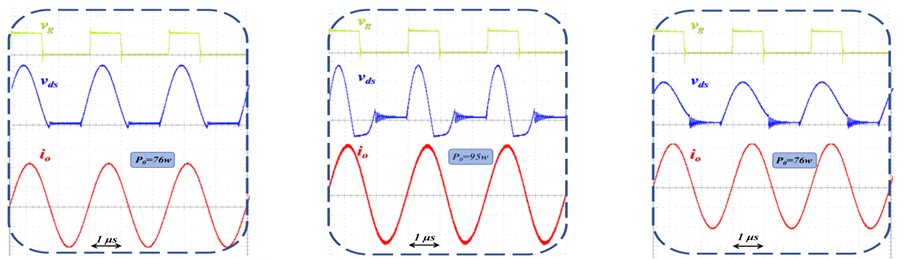}
    \caption{Waveform at different \( k \): \( v_{g} \) (5V/div), \( v_{ds} \) (50V/div), \( i_{0} \) (2A/div).
(a) \( k \) = 0.04. (b) \( k \) = 0.05. (c) \( k \) = 0.07.}
    \label{fig7}
\end{figure*}
The system employs a GaN-based transistor (GS66508B). It is important to note that a constant switch junction capacitance of 200 pF is taken into account and is to be incorporated into \( C_{1} \). Fig.\ \ref{fig7} demonstrates an experimental IPT system. 
\begin{figure}[htbp]
    \centering
    \includegraphics[width=0.9\linewidth]{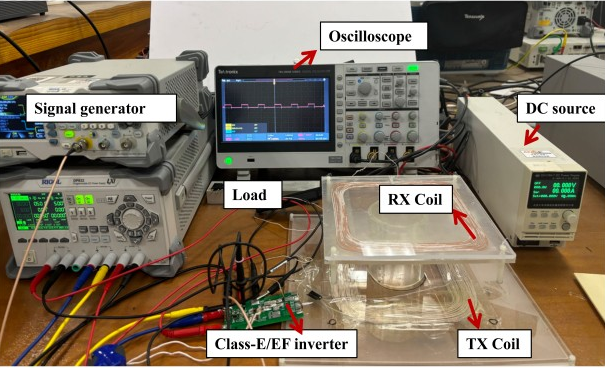}
    \caption{Experimental setup.}
    \label{fig6}
\end{figure}

Fig.\ \ref{fig7} graphs typical waveforms for different coupling conditions. It furthermore displays the measured output power and efficiency. The power fluctuation rate $\beta$ is 15\% which successfully aligning with the predictions made by the model-based design.  It can also be observed that as the coupling coefficient increases, the efficiency gradually improves. Although partial ZVS is lost, an efficiency above 85\% is still maintained, which verifies the feasibility of the design.

\begin{figure}[htbp]\hspace{-28pt}
    \centering
    \includegraphics[width=1.1\linewidth]{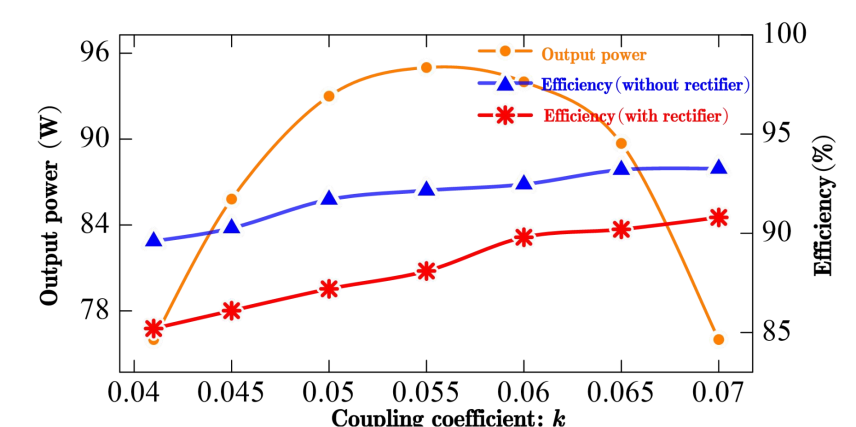}
    \caption{Output power and efficiency.}
    \label{fig5}
\end{figure}

\section{Conclusions}
This work solves the shortcomings of load-independent Class E/EF designs when subjected to low coupling variations. A detuned design on the secondary side can liberate global variables. Analysis with an expanded impedance model identifies the optimal design points. Experimental results indicate that under low coupling conditions, the power fluctuation rate  remains at 15\%, with efficiency maintained within the range of 86--91\% as the coupling coefficient \( k \) varies from 0.04 to 0.07.

	\bibliographystyle{IEEEtran}
	\bibliography{IEEEabrv,name}
	
	\vfill
\vfill
\end{document}